\documentclass[twocolumn,showpacs,preprintnumbers,amsmath,amssymb]{revtex4}
\usepackage{amsmath}
\usepackage{bm}
\usepackage{epsfig}
\usepackage[colorlinks=true,linkcolor=blue]{hyperref}

\newcommand{\be}{\begin{equation}}
\newcommand{\ee}{\end{equation}}

\newcommand{\bea}{\begin{eqnarray}}
\newcommand{\eea}{\end{eqnarray}}

\newcommand{\al}{\alpha}
\newcommand{\bt}{\beta}
\newcommand{\gm}{\gamma}

\newcommand{\dl}{\delta}
\newcommand{\Dl}{\Delta}

\newcommand{\et}{\eta}

\newcommand{\lm}{\lambda}
\newcommand{\ks}{\xi}
\newcommand{\rh}{\rho}

\newcommand{\ta}{\tau}

\newcommand{\om}{\omega}
\newcommand{\Om}{\Omega}

\newcommand{\Rarrow}{\Rightarrow}

\newcommand{\nn}{\nonumber}

\newcommand{\varep}{\varepsilon}

\begin{document}

\title {Imprints of cosmic strings on the cosmological GW background}

\author{K Kleidis$^{1,2}$, D B Papadopoulos$^1$, E Verdaguer$^3$ and L Vlahos$^1$}

\affiliation{$^1$Department of Physics, Aristotle University of
Thessaloniki, 54124 Thessaloniki, Greece}

\affiliation {$^2$Department of Civil Engineering, Technological
Education Institute of Serres, 62124 Serres, Greece}

\affiliation{$^3$Departament de Fisica Fonamental and Institut de
Ciences del Cosmos, Universitat de Barcelona, Av. Diagonal 647,
E-08028 Barcelona, Spain}


\date{\today}

\begin{abstract}
The equation which governs the temporal evolution of a
gravitational wave (GW) in curved space-time can be treated as the
Schr\"{o}dinger equation for a particle moving in the presence of
an {\em effective potential}. When GWs propagate in an expanding
Universe with {\em constant} effective potential, there is a {\em
critical value} $(k_c)$ of the comoving wave-number which
discriminates the metric perturbations into oscillating $(k >
k_c)$ and non-oscillating $(k < k_c)$ modes. As a consequence, if
the non-oscillatory modes are outside the horizon they do not {\em
freeze out}. The effective potential is reduced to a non-vanishing
constant in a cosmological model which is driven by a {\em
two-component fluid}, consisting of radiation (dominant) and
cosmic strings (sub-dominant). It is known that the cosmological
evolution gradually results in the {\em scaling} of a
cosmic-string network and, therefore, after some time $(\Dl \ta)$
the Universe becomes radiation-dominated. The evolution of the
non-oscillatory GW modes during $\Dl \ta$ (while they were outside
the horizon), results in the {\em distortion} of the GW power
spectrum from what it is anticipated in a {\em pure}
radiation-model, at present-time frequencies in the range
$10^{-16} \; Hz < f \lesssim 10^5$ $Hz$.
\end{abstract}


\maketitle

\section{Introduction}

The so-called cosmological gravitational waves (CGW) represent
small-scale perturbations to the Universe metric tensor~\cite{1}.
Since gravity is the weakest of the four known forces, these
metric corrections decouple from the rest of the Universe at very
early times, presumably at the Planck epoch~\cite{2}. Their
subsequent propagation is governed by the space-time curvature,
encapsulating in the field equations the inherent coupling between
relic GWs and the Universe matter-content; the latter being
responsible for the background gravitational field~\cite{3}.

In this context, we consider the coupling between CGWs and cosmic
strings. They are one-dimensional objects that can be formed as
linear defects at a symmetry-breaking phase transition~\cite{4},
~\cite{5}. If they exist, they may help us to explain some of the
large-scale structures seen in the Universe today, such as
gravitational lenses~\cite{6}. They may also serve as {\em seeds}
for density perturbations~\cite{7}, \cite{8}, as well as potential
sources of relic gravitational radiation~\cite{9}.

In the present article we explore another possibility: A fluid of
cosmic strings could be responsible for the constancy of the
effective potential in the equation which drives the temporal
evolution of a CGW in an expanding Universe. As we find out, a
constant effective potential leads to a critical comoving
wave-number $(k_c)$, which discriminates the metric fluctuations
into oscillating modes $(k > k_c)$ and non-oscillatory $(k < k_c)$
ones. As long as the latter lie outside the horizon, they do not
freeze out, resulting in the departure of the inflationary GW
power-spectrum from scale-invariance. This would be the case, if
there is a short period after inflation where the cosmological
fluid is made out of radiation and a sub-dominant component of
cosmic strings. As regards the space-time geometry itself, the
spatially flat Friedman - Robertson - Walker (FRW) model appears
to interpret adequately both the observational data related to the
known thermal history of the Universe and the theoretical approach
to cosmic string configurations~\cite{4}. Consequently, we will
assume our cosmological background to be a spatially flat FRW
model.

This Paper is organized as follows: In Section II we summarize the
theory of CGWs in curved space-time. In Section III we demonstrate
that, in a radiation model {\em contaminated} by a fraction of
cosmic strings, the effective potential in the equation which
governs the temporal evolution of a CGW in curved space-time is
constant. In Section IV we explore the characteristics of a
potential contribution of cosmic strings to the evolution of the
Universe and in Section V we study the propagation of the
non-oscillatory GW modes during this stage. We find that, if the
Universe evolution includes a {\em radiation-plus-strings} stage,
then, although it could last only for a short period of time, its
presence would lead to a distortion of the stochastic GW
background from what it is anticipated in a pure radiation model,
at present-time frequencies in the range $10^{-16} \; Hz < f
\lesssim 10^5 \; Hz$.

\section{Gravitational waves in curved space-time}

The far-field propagation of a weak CGW $(\vert h_{\mu \nu} \vert
\ll 1)$ in a curved, non-vacuum space-time is determined by the
differential equations~\cite{10} \be h_{\mu \nu ; \al}^{\; ; \al}
- 2 {\cal R}_{\al \mu \nu \bt} h^{\al \bt} = 0 \ee under the gauge
choice \be \left ( h^{\al \bt} - \frac{1}{2} g^{\al \bt}
h_{\mu}^{\mu} \right )_{; \bt} = 0 \ee which brings the linearized
Einstein equations into the form (1). In Eqs (1) and (2), Greek
indices refer to the four-dimensional space-time, ${\cal R}_{\al
\mu \nu \bt}$ is the Riemann curvature tensor of the background
metric, $h_{\mu}^{\mu}$ is the trace of $h_{\mu \nu}$ and the
semicolon denotes covariant derivative.

In the system of units where $c = 1$, a linearly-polarized, plane
GW propagating in a spatially flat FRW cosmological model, is
defined as~\cite{9} \be ds^2 = R^2 (\ta) \left [ d \ta^2 -
(\dl_{ij} + h_{ij} ) dx^i dx^j \right ] \ee where $\ta$ is the
conformal-time coordinate, Latin indices refer to the
three-dimensional spatial section and $\dl_{ij}$ is the Kronecker
symbol. The dimensionless scale factor $R(\ta)$ is a solution to
the Friedmann equation \be \left ( \frac{R^{\prime}}{R^2} \right
)^2 = \frac{8 \pi G}{3} \: \rh (\ta) \ee (where, the prime denotes
differentiation with respect to $\ta$ and $G$ is Newton's
constant), with matter-content in the form of a perfect fluid,
$T_{\mu \nu} = diag (\rho, - p, -p, -p)$, which obeys the
conservation law \be \rh^{\prime} + 3 {R^{\prime} \over R} \; (
\rh + p ) = 0 \ee and the equation of state \be p = \left ( {m
\over 3} - 1 \right ) \; \rh \ee where $\rh(\ta)$ and $p(\ta)$
represent the mass-density and the pressure, respectively.

The linear equation of state (6) covers most of the
matter-components considered to drive the evolution of the
Universe~\cite{11} - \cite{13}, such as quantum vacuum $(m = 0)$,
a network of domain walls $(m = 1)$, a gas of cosmic strings $(m =
2)$, dust $(m = 3)$, radiation $(m = 4)$ and Zel'dovich
ultra-stiff matter $(m = 6)$. For each component, the continuity
equation (5) yields \be \rh = {M_m \over R^m} \ee where $M_m$ is
an integration constant, associated to the {\em initial
mass-density} of the $m-$th component. Provided that the various
components do not interact with each other, a mixture of them
obeys~\cite{11} \be \rh = \sum_m {M_m \over R^m} \ee where, now,
Eq (5) holds for each matter-constituent separately.

In the case of an one component fluid, the Friedman equation (4)
reads \be R^{ {m \over 2} - 2} \: R^{\prime} = ( {8 \pi G \over 3}
M_m)^{1/2} \ee and, for every type of matter-content other than
cosmic strings $(m \neq 2)$, it results in \be R (\et) = ( {\et
\over \et_m } )^{{2 \over m - 2}} \ee where, the time-parameter
$\et$ is linearly related to the corresponding conformal one, by
$\et = {m - 2 \over 2} \: \ta$ and we have set $\et_m = ({8 \pi G
\over 3} M_m)^{-1/2} $. Notice that, for $m = 0$ (De Sitter
inflation) and $0 < \ta < \infty$, we obtain $- \infty < \et < 0$.

The general solution to Eq (1) in the curved space-time (3) is a
linear superposition of plane-wave modes \be h_{ij} (\ta ,
\vec{x}) = \frac{h_k (\ta)} {R (\ta)} \: \varep_{ij} \: e^{\imath
k_j x^j} \ee where $h_k (\ta)$ is the time-dependent part of the
mode denoted by $k$ and $\varep_{ij}$ is the polarization tensor,
depending only on the direction of the {\em comoving} wave-vector
$k_j$. Accordingly, for a fixed wave-number $k^2 = \sum k_j^2$,
the time-dependent part of the corresponding GW mode satisfies the
second-order differential equation~\cite{14}, \cite{15} \be
h_k^{\prime \prime} (\ta) + \left ( k^2 - \frac{R^{\prime
\prime}}{R} \right ) \: h_k (\ta) = 0 \ee Eq (12) can be treated
as the Schr\"{o}dinger equation for a particle moving in the
presence of the {\em effective potential} \be V_{eff} = {R^{\prime
\prime} \over R} \ee and, in a cosmological model of the form
(10), is written in the form \be h^{\prime \prime} (\et) + \left
\{ k_m^2 - 2 \left [ {4 - m \over (m - 2)^2} \right ] \: {1 \over
\et^2} \right \} \: h (\et) = 0 \ee yielding \be h_m (k_m , \et) =
\sqrt {\et} \: \left [ c_1 \: H_{\vert \nu \vert}^{(1)} (k_m ,
\et) + c_2 \: H_{\vert \nu \vert}^{(2)} (k_m , \et) \right ] \ee
where, now, a prime denotes derivative with respect to $\et$,
$c_1$ and $c_2$ are arbitrary constants to be determined by the
initial conditions and $k_m = {2 \over m - 2} \: k$, so that $k_m
\et = k \ta$. Finally, $H_{\vert \nu \vert}^{(1)}$ and $H_{\vert
\nu \vert}^{(2)}$ are the Hankel functions of the first and the
second kind, of order~\cite{16} \be \vert \nu \vert = {1 \over 2}
\left \vert {m - 6 \over m - 2} \right \vert \ee Therefore,
different types of matter-content (reflecting different periods in
the evolution of the Universe) admit different Hankel functions
(see also~\cite{17}).

\section{Constancy of the effective potential}

\subsection{Implications on CGWs' propagation}

A case of particular interest, involved in the time-evolution of a
primordial GW, is when the effective potential (13) is {\em
constant} for every $\ta$, namely \be {R^{\prime \prime} \over R}
= {8 \pi G \over 3} M \ee where $M$ is a non-negative constant of
dimensions $L^{-4}$. In this case, Eq (12) is written in the form
\be h_k^{\prime \prime} (\ta) + \om^2 \: h_k (\ta) = 0 \ee where
\be \om^2 = k^2 - {8 \pi G \over 3} M \ee is the (constant)
frequency of the wave. According to Eq (19), a {\em critical
value} of the comoving wave-number arises, through the condition
\be \om^2 \gtrless 0 \, \Leftrightarrow \, k \gtrless k_c =
\sqrt{{8 \pi G \over 3} M } \ee This critical value discriminates
the primordial GWs in modes with $k > k_c$, which oscillate for
every $\ta$, \be h_{k > k_c} (\ta) \sim e^{\imath \sqrt {k^2 -
k_c^2} \: \ta} \ee and modes with $k < k_c$, which grow
exponentially for every $\ta$, \be h_{k < k_c} (\ta) \sim e^{\sqrt
{k_c^2 - k^2} \: \ta} \ee (the exponentially decaying solutions
are neglected).

\subsection{Cosmological models of constant effective potential}

Now, the question arises on whether there exists a spatially flat
FRW cosmological model in which the effective potential is
constant. To answer this question, we set ${\cal H} = {R^{\prime}
\over R}$. Accordingly, $V_{eff}$ is written in the form \be
{R^{\prime \prime} \over R} = {\cal H}^{\prime} + {\cal H}^2 \ee
Upon consideration of Eqs (4) and (23), Eq (17) results in the
ordinary differential equation \be \rh^{\prime} + 4 \rh
{R^{\prime} \over R} = 2 M {R^{\prime} \over R^3} \ee which admits
the solution \be \rh (\ta) = {C \over R^4} + { M \over R^2} \ee
where $C$ is an arbitrary integration constant of dimensions
$L^{-4}$. In comparison to Eqs (7) and (8), we distinguish the
following cases:

{\bf (i)} $M = 0$ and $C = 0$: This case corresponds to {\em
vacuum and flat space-time} and it will not be considered further.

{\bf (ii)} $M = 0$ and $C \neq 0$: This choice results in the {\em
radiation-dominated Universe} \be \rh (\ta) = {C \over R^4} \; \;
, \; \; V_{eff} = 0 \ee where the critical wave-number vanishes
$(k_c = 0)$.

{\bf (iii)} $M \neq 0$ and $C = 0$: Hence, \be \rh (\ta) = {M
\over R^2} \; \; , \; \; V_{eff} = {8 \pi G \over 3} M \ee which
corresponds to a {\em string-dominated Universe}~\cite{18}. It is
worth noting that the constant $M$ appearing in the effective
potential (17) is, in fact, the initial mass-density of the
strings, $M_2$ [clf Eq (7)]. A string-dominated Universe does not
seem likely~\cite{19}, \cite{20} and, therefore, this case is of
no particular interest.

{\bf (iv)} Finally, if both $C$ and $M$ differ from zero, then,
the function $\rh (\ta)$ consists of two parts: One evolving as
$R^{-4}$ and the other as $R^{-2}$. By analogy to Eq (8), this
type of matter-content can be met in a cosmological model filled
with relativistic particles (radiation) and a fluid of cosmic
strings, without interacting with each other, as it should be the
case shortly after the {\em dynamic friction} between
them~\cite{4} became unimportant~\cite{5}. Therefore, in this
case, \be \rh = {M_4 \over R^4} + {M_2 \over R^2} \; \; , \; \;
V_{eff} = {8 \pi G \over 3} M_2 \ee Once again, the constant $M$,
appearing in the effective potential, is associated to the initial
amount of strings in the mixture. It appears that, whenever the
effective potential acquires a non-zero constant value, this value
always involves the initial density of a cosmic-string gas.

We conclude that, in the presence of cosmic strings the effective
potential is reduced to a non-vanishing constant and, therefore,
oscillation of the metric perturbations is possible only if their
comoving wave-number is larger than a critical value, depending on
the mass-density of the linear defects \be k > k_c = \sqrt{{8 \pi
G \over 3} M_2} \ee In other words, a cosmic-string network
discriminates the primordial GWs predicted by inflation into
oscillating and non-oscillating modes, something that should be
reflected in the power-spectrum of the stochastic GW background.
We shall attempt to illustrate how, in a realistic setting.

\section{A Universe with cosmic strings}

The presence of cosmic strings in a unified gauge theory is purely
a question of topology. The simplest SO(10) model, for example,
predicts strings~\cite{21}. Many {\em superstring-inspired} models
also result in the formation of linear topological
defects~\cite{22}, \cite{23}. Cosmic strings are formed at a
symmetry-breaking phase transition, within the radiation-dominated
epoch \be R (\ta) = {\ta \over \ta_{cr}} \ee where $\ta_{cr}$ is
the time at which the Universe acquires the {\em critical
temperature} below which the strings are formed and we have
normalized $R (\ta_{cr})$ to unity.

In particular, after inflation (and reheating) the Universe enters
in an {\em early-radiation} epoch~\cite{24}, during which the
background temperature drops monotonically $(T \sim R^{-1})$. For
$\ta \geq \ta_{cr}$, this cooling process results in the breaking
of a fundamental U(1) local gauge symmetry, leading to the
formation of linear defects (for a detailed analysis see~\cite{4}
and/or~\cite{5}).

By the time the cosmic strings are formed, they are moving in a
very dense environment and, hence, their motion is heavily damped
due to string-particle scattering~\cite{25} - \cite{28}. This
friction becomes subdominant to expansion damping at~\cite{25} \be
\ta_* = \frac{1}{\sqrt {G \mu}} \: \ta_{cr} \ee where, $\mu$ is
the mass per unit length of the linear defect. For $\ta \geq
\ta_*$, the motion of long cosmic strings can be considered
essentially independent of anything else in the Universe and soon
they acquire relativistic velocities. Therefore, we may consider
that, after $\ta_*$ the evolution of the Universe is driven by a
two-component fluid, consisting of relativistic particles
(dominant) and cosmic strings (sub-dominant). Consequently, Eq
(28) holds and $\ta_*$ marks the beginning of a {\em
radiation-plus-strings} stage. During this stage, the Friedman
equation (4) yields \be R (\ta) = \sqrt{ \frac{M_4}{M_2}} \; \sinh
\sqrt{ \frac{8 \pi G}{3} M_2} \; \ta \ee Nevertheless, the scale
factor (32) can drive the Universe expansion only for a short
period of time after $\ta_*$, since cosmic strings should (at any
time) be a small proportion of the Universe energy-content. This
means that the equation of state considered in (28) should have
validity only for a limited time-period, otherwise cosmic strings
would eventually dominate the overall energy-density~\cite{18}.

In fact, a radiation-plus-strings stage (if ever existed) does not
last very long. Numerical simulations~\cite{29} - \cite{32}
suggest that, after the friction becomes unimportant, the
production of loops smaller than the Hubble radius gradually
results in the {\em scaling} of the long-string network.
Accordingly, the linear defects form a self-similar configuration,
the density of which, eventually, behaves as $R^{-4}$~\cite{4}. In
this way, apart from small statistical fluctuations, at some time
$\ta_{sc} > \ta_*$ the Universe re-enters in the ({\em late})
radiation era \be R (\ta) = R_{sc} \: {\ta \over \ta_{sc}} \ee
before it can become string-dominated. The duration $(\Dl \ta =
\ta_{sc} - \ta_*)$ of the radiation-plus-strings stage is quite
uncertain, mostly due to the fact that numerical simulations can
be run for relatively limited times. For example, the longest run
of~\cite{32} suggests that $\ta_{sc} \simeq 4.24 \: \ta_*$
(corresponding to a factor of $18$ in terms of the physical time),
while~\cite{31} raise this value to $\ta_{sc} \simeq 6.48 \:
\ta_*$ $(t_{sc} \simeq 42 \: t_*)$.

In what follows, we explore the evolution of GW modes with $k <
k_c$ through the radiation-plus-strings stage.

\section{CGWs in the presence of cosmic strings}

\subsection{Evolution of modes outside the horizon}

CGWs are produced by quantum fluctuations during inflation (e.g.
see~\cite{33}). Some of them escape from the visible Universe,
once their reduced physical wavelength $[\lm_{ph} = {\lm \over 2
\pi} R (\ta)]$ becomes larger than the (constant) inflationary
horizon $[\ell_H = H_{dS}^{-1}$, $H_{dS}$ being the Hubble
parameter of the de Sitter space]. Eventually, every CGW with $k
\leq k_{max} = H_{dS} R_{dS}$ is {\em exiled} from the Hubble
sphere and {\em freezes out}, acquiring the constant
amplitude~\cite{34}, \cite{35} \be \al_k^2 = \left [ \frac{h_k
(\ta)}{R (\ta)} \right ]^2 = \frac{16}{\pi} \: \left (
\frac{H_{dS}}{m_{Pl}} \right )^2 \: k^{-3} \ee where $m_{Pl} =
G^{-1/2}$ is the Planck mass. After inflation, i.e. within the
subsequent radiation epoch, analytic solutions for $\al_k (\ta)$
can be expressed in terms of the Bessel function $J_{1 \over 2} (k
\ta)$ \be \al_k (\ta) = 2 c_1 \: \al_k \: {\sqrt {\ta} \over R
(\ta)} \: J_{1 \over 2} (k \ta) \sim \al_k \: {\sin k \ta \over k
\ta} \ee [clf Eq (15) for $c_1 = c_2$]. Accordingly, when $k \ta
\ll 1$, the perturbation's amplitude evolves slowly and is
approximately constant. Once $k \ta \approx 1$, the amplitude
decays away rapidly before entering in an oscillatory phase with
slowly decreasing amplitude, when $k \ta \gg 1$. Physically, this
corresponds to a mode that is (almost) frozen beyond the horizon,
until its physical wavelength becomes comparable to the Hubble
radius, at which point it enters in the visible Universe (e.g.
see~\cite{36}).

In other words, as the Universe expands, a fraction of the modes
that lie beyond the horizon re-enters inside the Hubble sphere. At
the time of re-entry their amplitude is given by Eq (34), while,
afterwards, they begin oscillating. The $k$-dependence of their
amplitude implies a {\em scale-invariant}
power-spectrum~\cite{37}.

However, if the cosmological evolution includes a
radiation-plus-strings stage, then, during this stage, the
effective potential is a non-vanishing constant. In other words,
$k_c \neq 0$ and the equation which governs the temporal evolution
of the GW modes with $k < k_c$ does not admit the solution (35),
but Eq (22). As a consequence, even if they lie outside the
horizon, {\em these modes do not freeze out}.

At the beginning of the radiation-plus-strings stage, the GW modes
that fit inside the visible Universe obey the condition \be
\lm_{ph} (\ta_*) \leq \ell_H (\ta_*) \; \Rarrow \; k \geq H
(\ta_*) R (\ta_*) = {1 \over \ta_*} \ee while, modes of $k < k_* =
\ta_*^{-1}$ lie outside the horizon. In order to examine whether
$k_c \gtrless k_*$ we need to determine the initial mass-density
of the linear defects, since, by definition, \be M_2 = \rh_{str}
(\ta_*) R^2 (\ta_*) \ee Let us consider a network of cosmic
strings characterized by a {\em correlation length} $\ks (\ta)$.
This may be defined as the length such that the mass within a
typical volume $\ks^3$, is $\mu \ks$~\cite{4}. In this case, at
$\ta = \ta_*$, the cosmic strings contribute to the Universe
matter-content a mean density \be \rh_{str} (\ta_*) = {\mu \over
\ks^2 (\ta_*)} = \gm_*^2 \: {\mu \over \ell_H^2 (\ta_*)} \ee where
$\gm_*$ is a numerical constant of the order of unity,
representing the number of correlation lengths inside the horizon
at $\ta_*$. Accordingly, \be M_2 = \mu \: \left ( {\gm_* \over
\ta_*} \right )^2 \ee and hence \be k_c = \sqrt {8 \pi \over 3} \:
\sqrt {G \mu} \: \: {\gm_* \over \ta_*} \ee For GUT-scale strings
we have $(G \mu) \sim 10^{-6}$ and $\gm_* \simeq 7$ (e.g.
see~\cite{4}), so that $k_c \simeq 2 \times 10^{-2} \: k_* \ll
k_*$. In other words, at the beginning of the
radiation-plus-strings stage, the GW modes of $k < k_c$ do not fit
inside the horizon.

On the other hand, for $\ta_* < \ta \leq \ta_{sc}$, the condition
of fitting inside the Hubble sphere is written in the form \be {k
\over k_c} \geq \coth \sqrt{{8 \pi G \over 3} M_2} \: \ta \ee
Since the {\em hyperbolic cotangent} on the rhs is larger than
unity for every $\ta$, Eq (41) suggests that the GW modes of
comoving wave-numbers $k < k_c$ remain outside the horizon during
the whole radiation-plus-strings stage.

Nevertheless, by virtue of Eq (22), for $\ta_* < \ta \leq
\ta_{sc}$ their amplitude continues to evolve as \be \al_{k < k_c}
(\ta > \ta_*) = {4 \over \sqrt {\pi}} \: \left ( {H_{dS} \over
m_{Pl}} \right ) \: {1 \over k^{3/2}} \: \left [ {R (\ta_*) \over
R (\ta)} \right ] \: e^{\sqrt {k_c^2 - k^2} \: (\ta - \ta_*)} \ee
[clf Eqs (22) and (34)]. This behavior ends at $\ta_{sc}$, when
the scaling of the long-string network is completed and the
Universe re-enters in the (late) radiation era. For $\ta >
\ta_{sc}$ the GW modes of $k < k_c$ are no longer influenced by
the radiation-plus-strings stage and therefore, just like the rest
of the metric perturbations outside the horizon, {\em (re)freeze
out}. As a consequence, their amplitude acquires the constant
value \be \al_{k < k_c} = {4 \over \sqrt {\pi}} \: \left ( {H_{dS}
\over m_{Pl}} \right ) \: {1 \over k^{3/2}} \: \left [ {R (\ta_*)
\over R (\ta_{sc})} \right ] \: e^{\sqrt {k_c^2 - k^2} \: \Dl \ta}
\ee

\subsection{The distorted power-spectrum}

Within the late-radiation era these modes remain frozen until the
time $\ta_c$. At that time the mode $k_c$ enters inside the
visible Universe, since its physical wavelength $(\lm_{c_{ph}}
\sim \ta_c)$ becomes smaller than the corresponding Hubble radius
$(\ell_H \sim \ta_c^2)$. In accordance, for $\ta > \ta_c$, GW
modes of $k < k_c$ also enter inside the Hubble sphere. After
entering inside the horizon, the GW modes under consideration
begin oscillating, thus producing a part of the power-spectrum we
observe today (or at some time in the future). However, since they
have experienced the influence of the radiation-plus-strings ({\it
rps}) stage, their amplitude is no longer given by Eq (34), but by
Eq (43), thus resulting in the distortion of the GW power-spectrum
$(P_k^2 \sim k^3 \: \al_k^2)$, from what it is anticipated by {\em
pure-radiation} ({\it rad}), at comoving wave-numbers $k < k_c$.
Namely, \be P_{k < k_c}^{rps} = P_{k < k_c}^{rad} \: {R (\ta_*)
\over R (\ta_{sc})} \: e^{\sqrt {k_c^2 - k^2} \: \Dl \ta} \ee
which, upon consideration of Eq (32), is written in the form \be
{P_{k < k_c}^{rps} \over P_{k < k_c}^{rad}} = {2 \: e^{(1 + \sqrt
{1 - x^2}) \: k_c \: \Dl \ta} \over [\coth (k_c \ta_*) + 1] \:
e^{2 k_c \Dl \ta} - [\coth (k_c \ta_*) - 1]} \ee where, we have
set $0 < {k \over k_c} = x = {f \over f_c} < 1$ and $f$ is the
frequency attributed to the GW mode denoted by $k$. According to
Eq (40), $\coth (k_c \ta_*) \simeq 5$ and Eq (45) results in \be
{P_{k < k_c}^{rps} \over P_{k < k_c}^{rad}} = {e^{(1 + \sqrt {1 -
x^2}) \: k_c \: \Dl \ta} \over 3 \: e^{2 k_c \Dl \ta} - 2} \ee
Clearly, for $\Dl \ta = 0$ (i.e. in the absence of the
radiation-plus-strings stage) $P_{k < k_c}^{rps} = P_{k <
k_c}^{rad}$, while, for $\Dl \ta \neq 0$ the inflationary-GW
power-spectrum is no longer scale-invariant.

The {\em spectral function} $\Om_{gw}$, appropriate to describe
the intensity of a stochastic GW background~\cite{38}, is related
to the power-spectrum as $\Om_{gw} \sim P_k^2$ (e.g.
see~\cite{39}). Therefore, Eq (46) yields \be {\Om_{gw}^{rps} (f <
f_c) \over \Om_{gw}^{rad} (f < f_c)} = {e^{2 (1 + \sqrt {1 - x^2})
\: k_c \: \Dl \ta} \over \left ( 3 \: e^{2 k_c \Dl \ta} - 2 \right
)^2} \ee Notice that, for every $0 \leq x \leq 1$, we have
$\Om_{gw}^{rps} \leq \Om_{gw}^{rad}$, with the equality being
valid only for $\Dl \ta = 0$. In other words, the involvement of a
radiation-plus-strings stage in the evolution of the Universe
reduces the stochastic GW intensity to lower levels than those
expected by pure radiation. To give some numbers, we take into
account the numerical results of~\cite{31}, as well as those
of~\cite{32}. Accordingly, a reasonable estimate on the duration
of radiation-plus-strings stage would be $\ta_{sc} = 5.5 \: \ta_*$
and therefore, $k_c \Dl \ta \simeq 9 \times 10^{-2}$. In this
case, Eq (47) is written in the form \be {\Om_{gw}^{rps} (f < f_c)
\over \Om_{gw}^{rad} (f < f_c)} \simeq 0.47 \times e^{0.18 \sqrt
{1 - x^2}} \ee from which it becomes evident that, for $f < f_c$,
the value of $\Om_{gw}$ is no longer $8 \times 10^{-14}$, as it is
predicted by pure radiation~\cite{2}, \cite{9}, but rather \be
\Om_{gw}^{rps} \simeq 0.5 \: \Om_{gw}^{rad} \simeq 4 \times
10^{-14}. \ee Such a distortion reflects a change in the
distribution of the GW energy-density among the various frequency
intervals, probably due to the coupling between metric
perturbations and cosmic strings.

The question that arises now is, whether these results are
observable by the detectors currently available. To answer this
question, we should determine explicitly both $f_c$ (the critical
frequency) and $t_c$ (the physical time at which the GW modes of
$f < f_c$ begin entering inside the horizon). In what follows, $c
\neq 1$.

\begin{figure}[h!]
\centerline{\mbox {\epsfxsize=9.cm \epsfysize=7.cm
\epsfbox{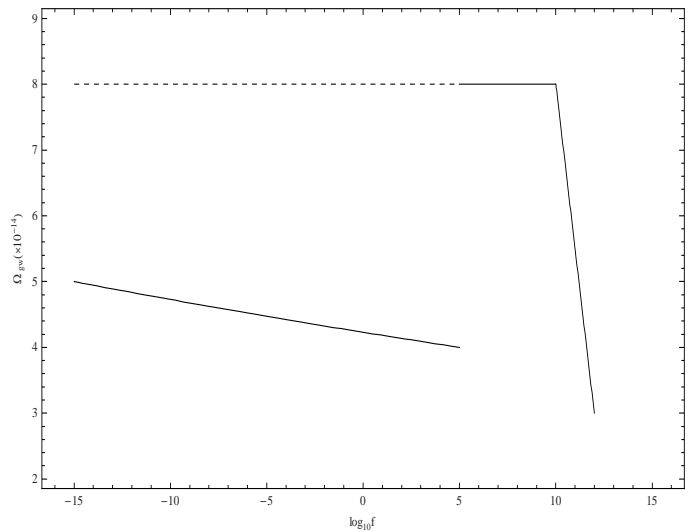}}} \caption{The stochastic GW
background from inflation at frequencies within the radiation era,
$f \gtrsim 10^{-16} \; Hz$ (dashed line), under the influence of a
radiation-plus-strings stage produced by GUT-scale cosmic strings
(solid line).}
\end{figure}

During the early-radiation epoch, the {\em physical time} is
defined as \be t = \int R(\ta) d \ta \; \Rarrow \; t = {\ta^2
\over 2 \ta_{cr}} \ee With the aid of Eqs (31) and (50), Eq (40)
is written in the form \be k_c = \sqrt {2 \pi \over 3} \: \left (
{G \mu \over c^2} \right ) \: {\gm_* \over c \: t_{cr}} \ee and
therefore \be f_c = {1 \over \sqrt{6 \pi}} \: \left ( {G \mu \over
c^2} \right ) \: {\gm_* \over t_{cr}} \ee The first of the GW
modes under consideration which enters inside the visible Universe
is the one with the shortest comoving wavelength $(\lm_c)$, i.e.
the one with the largest frequency $(f_c)$. In terms of the
physical time, this process begins at $t_c$, at which
$\lm_{c_{ph}} (t_c) \leq \ell_H (t_c)$.

Within the late-radiation era, the physical time is defined as \be
t = \int R(\ta) d \ta \; \Rarrow \; t = R_{sc} {\ta^2 \over 2
\ta_{sc}} \ee and therefore \be t_c \: \gtrsim \: {3 \pi \over 11
\gm_*^2} \: \left ( {G \mu \over c^2} \right )^{-1} t_* \: = \: {3
\pi \over 11 \gm_*^2} \: \left ( {G \mu \over c^2} \right )^{-2}
t_{cr} \ee where we have used Eq (31) and the fact that, in an
expanding Universe, \be R_{sc} > R (t_*) = \left ( {G \mu \over
c^2} \right )^{-1/2}. \ee Within the Hubble sphere the GW modes of
$\lm > \lm_c$ correspond to CGWs of frequencies $f < f_c$.
Extrapolation of this result into the present epoch $(t_{pr}
\simeq 13.7 \times 10^9 \; y)$, suggests that at frequencies \bea
f^{pr} < f_c^{pr} & = & f_c \: \left ( {t_c \over t_{rec}} \right
)^{1/2} \: \left ( {t_{rec} \over t_{pr}} \right )^{2/3} \: \Rarrow \nn \\
f^{pr} < f_c^{pr} & = & {1 \over \sqrt{22}} \: \left ( {t_{cr}
\over t_{rec}} \right )^{1/2} \: \left ( {t_{rec} \over t_{pr}}
\right )^{2/3} \: {1 \over t_{cr}} \eea (where $t_{rec} = 1.2
\times 10^{13} \; sec$ is the {\em recombination time}) the
inflationary-GW power-spectrum is distorted, departing from
scale-invariance.

We note that $f_c^{pr}$ depends only on the (physical) time at
which the cosmic strings are formed. These linear defects may have
been formed at a grand unification (GUT) transition or,
conceivably, much later, at the electro-weak transition or
somewhere in between~\cite{4}. For GUT-scale strings, $t_{cr} \sim
10^{-31} \; sec$~\cite{5} and therefore $f_c^{pr} \simeq 1.5
\times 10^5 \; Hz$. Clearly, this value is far outside of the
range where both the ground-based and the space-based laser
interferometers may operate. A GW of this frequency could be
detected only by a system of {\em coupled super-conducting
microwave cavities}~\cite{40}, \cite{41}.

However, one should have in mind that, this is only the {\em upper
bound} of the distorted GW power-spectrum. In fact, if cosmic
strings contribute to the evolution of the Universe, the GW
power-spectrum will decline from what it is anticipated by pure
radiation {\em at every present-time frequency} in the range
$10^{-16} \; Hz < f \lesssim f_c^{pr}$ (see Fig. 1). The lower
bound of this range arises from the GWs that began entering inside
the horizon after the Universe has become
matter-dominated~\cite{39}.

On the other hand, for electro-weak-scale strings, $t_{cr} \sim
10^{-11} \; sec$~\cite{5} and hence, $f_c^{pr} \simeq 1.5 \times
10^{-5} \; Hz$, while, for cosmic strings created at some time in
between the GUT- and the electro-weak-symmetry breaking (e.g.
$t_{cr} \sim 10^{-21} \; sec$), we obtain $f_c^{pr} \simeq 1.5 \;
Hz$. Therefore, a potential detection of CGWs, among other things,
would give us valuable information on the epoch (and therefore on
the physical mechanism, as well) at which the cosmic strings were
formed.

\section{Conclusions}

The equation which governs the temporal evolution of a CGW in a
Friedman Universe can be treated as the Schr\"{o}dinger equation
for a particle moving in the presence of the effective potential
$V_{eff} = R^{\prime \prime} / R$. In the present article we show
that, if there is a period where the effective potential is
constant, this would lead to a critical value $(k_c)$ in the
comoving wave-number of the metric fluctuations, discriminating
them into oscillating $(k > k_c)$ and non-oscillating $(k < k_c)$
modes. As a consequence, when the non-oscillatory modes lie
outside the horizon do not freeze out, something that should be
reflected in the inflationary-GW power-spectrum.

This property is met in a radiation model contaminated by a
fraction of cosmic strings. Therefore, if the cosmological
evolution includes a radiation-plus-strings stage, some of the
long-wavelength GW modes (although being outside the Hubble
sphere) continue to evolve. However, this stage (if ever existed)
does not last very long, since, the cosmological evolution
gradually results in the scaling of the cosmic-string network and,
after some time $(\Dl \ta)$, the Universe enters in the
late-radiation era.

In a radiation-dominated Universe the metric perturbations of $k <
k_c$ can enter the horizon, which now expands faster than their
physical wavelength. However, the evolution of the non-oscillatory
GW modes during $\Dl \ta$ (while they were outside the horizon)
has modified their amplitude and, therefore, oscillation of these
modes within the Hubble sphere, results in the distortion of the
scale-invariant GW power-spectrum at present-time frequencies in
the range $10^{-16} \; Hz < f \lesssim 10^5 \; Hz$.

\section*{Note added in proof}

The day after this article was accepted for publication, it came
to our attention a recent paper~\cite{42} dealing with the scaling
of a cosmic-string network in an updated numerical fashion.
According to it, the duration of a potential
radiation-plus-strings stage in terms of the conformal time (the
{\em dynamical range} - as it is referred to) is $\ta_{sc} = 17 \:
\ta_*$ (corresponding to a factor of $300$ in the physical time).
Adaption of this result would lead to {\em a more evident
distortion} of the inflationary GW spectrum, modifying Eqs (48)
and (49) to \be {\Om_{gw}^{rps} (f < f_c) \over \Om_{gw}^{rad} (f
< f_c)} \simeq 0.14 \times e^{0.64 \sqrt {1 - x^2}} \ee and \be
\Om_{gw}^{rps} \simeq 0.2 \: \Om_{gw}^{rad} \simeq 1.6 \times
10^{-14} \ee respectively. The authors would like to thank
Professor Mairi Sakellariadou for pointing that out.

\vspace{.3cm}

{\bf Acknowledgements:} This project was supported by the Greek
Ministry of Education, through the PYTHAGORAS Program and by the
Spanish Research Projects MEC FPA-2004-04582 and DURSI
2005SGR00082.

\end{document}